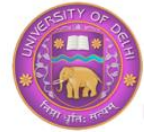

# Light curve modeling of eclipsing binaries towards the constellation of Carina

Aniruddha Dey[1], Sukanta Deb[1*], Subhash Kumar[1], Hrishabh Bhardwaj[1], Barnmoy Bhattacharya[1], Richa[1], Angad Sharma[1], Akshyata Chauhan[1], Neha Tiwari[1], Sharanjit Kaur[2], Suman Kumar[2], Abhishek[2]

sukantodeb@gmail.com

[1]Department of Physics, Acharya Narendra Dev College, Govindpuri, Kalkaji, New Delhi 110019, India

[2]Department of Computer Science, Acharya Narendra Dev College, Govindpuri, Kalkaji, New Delhi 110019, India

## ABSTRACT

We present a detailed *V*-band photometric light curve modeling of 30 eclipsing binaries using the data from Pietrukowicz et al. (2009) collected with the European Southern Observatory Very Large Telescope (ESO VLT) of diameter 8-m. The light curve of these 30 eclipsing binaries were selected out of 148 of them available in the database on the basis of complete phase coverage, regular and smooth phased light curve shapes. Eclipsing binaries play pivotal role in the direct measurement of astronomical distances more accurately simply from their geometry of light curves. The accurate value of Hubble constant ($H_0$) which measures the rate of expansion of the Universe heavily relies on extragalactic distance scale measurements. Classification of the selected binary stars in the sample were done, preliminarily on the basis of Fourier parameters in the $a_2$-$a_4$ plane and final classification was obtained from the Roche lobe geometry. Out of these 30 eclipsing binaries, only one was found to be detached binary system while the rest 29 of them belong to the contact binary systems. These contact binaries were further classified into the *A*-type and *W*-type based on their mass ratios. Since spectroscopic mass ratio measurements were not available for any of these binary stars, we determined the mass ratios through photometric light curve modeling with the aid of Wilson-Devinney code as implemented in PHOEBE. Various geometrical parameters and physical parameters of astrophysical importance viz., mass, radius and luminosity were obtained from the light curves of the selected stars.

## INTRODUCTION

Eclipsing Binaries are 'standard candles' which allow direct measurement of fundamental stellar parameters and distances precisely and accurately. The determination of size and structure of the entire galaxy can be obtained simply from the three dimensional distance distribution of these standard candles (18, 2, 22). Because of the accuracy and precision in the measurement of astronomical distances, they can be used to calibrate the zero point of the cosmic distance scale with an accuracy of 2% (22). They serve as an important astrophysical laboratory for confirming the existing theories about the origin, structure and evolution of stars and proxies for future hypothesis. Through combined photometric and spectroscopic analysis they accurately provide the temperatures and radii of the individual components of the binary stars (5).

Based on the Roche lobe geometry, a class of eclipsing binaries called the contact binaries exist in which both the components are clustered around the main sequence in the HR diagram. W UMa contact binaries, the most common type, are composed of stars with convective envelopes with spectral types late *F* to *K*. Lucy (1968) provided the first explanation of these binaries in terms of two stellar cores being surrounded by a common convective envelope at nearly constant temperature.(13) W UMa type binaries were subdivided into two groups, A-type and W-type systems depending on whether the more massive star or the less massive star was eclipsed at primary eclipse. The temperatures of the components of the W UMa type binaries are roughly equal since they share a common envelope having the same entropy (19). But Csizmadia and Klagyivik (2004) showed another subclass of W UMa contact binaries, called them B-type, in which the difference in temperature of components is more than 1000 K (3). These type of contact binaries are also referred as 'poor thermal contact binaries' (25).

There are many studies in the literature which have utilized eclipsing binary stars to determine the various Galactic and extragalactic distances. Using 10 eclipsing binaries, Harries, Hilditch, and Howarth (2003) determined the distance scale of the Small Magellanic Cloud (SMC) to be 18.89±0.14 mag (8). Later on, selecting 40 eclipsing binaries, Hilditch, Howarth, and Harries (2005) refined the distance modulus of the SMC to 18.91±0.10 mag from the fundamental parameters of these binaries (9). Recently, Graczyk et al. (2014) determined the distance to the SMC as (*m-M*)=18.965±0.025(*stat.*)±0.048(*sys.*) mag which corresponds to a distance of 62.1±1.9 kpc. Using 8 well-detached eclipsing binary systems from the Optical Gravitational Lensing Experiment (OGLE), (22) determined the distance modulus of the Large Magellanic Cloud (LMC) to (*m-M*)=18.493±0.19(*stat.*) ±0.008(*sys.*) mag which corresponds to a distance of 49.97±0.19(*stat.*) ±1.11(*sys.*) kpc. The accurate distance determination to the LMC plays a key role in constraining the value of the Hubble constant $H_0$ (22).

The various importance of eclipsing binaries in astrophysics has motivated us to investigate their properties from the available data obtained from ESO VLT. In section 1, we describe the data selected for the present study. In section 2, we use the Fourier parameters in the $a_2$-$a_4$ plane obtained from the Fourier cosine decomposition method in order to classify the stars in our sample for preliminary classification. Section 3 deals with the light curve modeling of these binaries using the Wilson-Devinney light curve modeling code as implemented in PHOEBE and in section 4, the various geometrical and physical parameters of the selected binary systems obtained from the light curve modeling are presented. Finally in section 5, a brief summary and conclusion of this study has been presented.

## THE DATA

The *V*-band light curve data for the present investigation were obtained from the observations carried out using an 8-m telescope with VIMOS at the Unit Telescope 3 (UT3) of the European Southern Observatory Very Large Telescope (ESO VLT) located at Paranal Observatory from April 9 to 12, 2009 (21). The aim of the observation was to search for variable objects in a deep (magnitude limit $V \sim 24.5$ mag) galactic field in the constellation of Carina (21). The observations resulted in the discovery of 348 variables among 50897 stars in the *V*-band magnitude range 15.4-24.5 mag, out of which 343 variables were new discoveries (21). VIMOS is an imager and multi-object spectrograph (Le Fèvre et al., 2003). Its field of view consists of 4 quadrants of about 7'x8' each, separated by a cross, 2'wide. The CCD size has 2048x2440 array of pixels with a pixel size of $0''.205$. Equatorial coordinates of the center of the field are RA(2000.0)=$10^h52^m56^s$ and Dec (2000.0)=-$61°28'15''$ or l = $289°.269$, b=$1°.783$ (21). The catalog contains variable star data with RA ($\alpha$), DEC ($\beta$), period (*P*), *V*-band magnitudes and their types. Most of the data for our analysis primarily belong to a class of contact binaries known as the W UMa type. These low mass eclipsing binaries consist of ellipsoidal components having orbital periods less than one day with continuously changing brightness (11, 20, 5). Their periods are usually in the range $0.2\ d<P<0.8\ d$. Both the components of these binary stars are of solar type main-sequence. They fill their Roche lobes sharing a common outer envelope at the inner Lagrangian point (28, 5). A characteristic property of their light curves is that they have rounded maxima. The proximity of the two stars causes their shapes to be strongly tidally distorted (28). Because of the possibility of complete phase coverage of light curves observed using a large telescope of diameter 8-m for only 4-days, these short period and distant binary stars provide an opportunity to study their various geometrical and physical parameters. Out of 148 eclipsing binaries identified in Pietrukowicz et al. (2009), 30 light curves were selected based on the accurate and complete phased light curves for further photometric analysis using PHysics Of Eclipsing BinariEs (PHOEBE) (23) . We attempted to recalculate the periods of all the 30 binary stars by using the minimization of entropy method as stated in Deb and Singh (2011) but these values differed very slightly from the original periods and did not affect the phased light curve shapes. We had used the recalculated periods in this work. The selected binary stars along with their equitorial coordinates, recalculated periods and *V*-band magnitude are listed in the Table 1.

**Table1. The 30 eclipsing binary stars used for the analayis.**

| Sl.no. | ID | RA(2000.0) | Dec(2000.0) | Period (d) | Primary minima (HJD 2400000+) | Amp (mag) | V (mag) |
|---|---|---|---|---|---|---|---|
| 1 | v028 | 10:51:53.53 | -61:34:11.2 | 0.2803 | 53470.163088 | 0.82 | 18.66 |
| 2 | v043 | 10:51:59.15 | -61:35:28.0 | 0.3791 | 53470.114974 | 0.44 | 18.70 |
| 3 | v056 | 10:52:02.44 | -61:22:47.6 | 0.3348 | 53470.240198 | 0.37 | 17.31 |
| 4 | v066 | 10:52:05.47 | -61:21:11.5 | 0.3104 | 53470.236936 | 0.18 | 17.80 |
| 5 | v073 | 10:52:08.47 | -61:33:34.7 | 0.4519 | 53470.184357 | 0.82 | 18.82 |
| 6 | v083 | 10:52:12.10 | -61:35:27.5 | 0.2995 | 53470.213162 | 1.20 | 21.10 |
| 7 | v084 | 10:52:12.54 | -61:24:53.5 | 0.2664 | 53470.079600 | 0.78 | 19.02 |
| 8 | v116* | 10:52:24.58 | -61:22:27.9 | 0.3012 | 53470.081083 | 0.34 | 19.17 |
| 9 | v143 | 10:52:32.92 | -61:29:33.8 | 0.3656 | 53470.243931 | 0.14 | 19.90 |
| 10 | v178 | 10:53:10.37 | -61:32:39.2 | 0.3967 | 53470.022917 | 0.70 | 20.05 |
| 11 | v198 | 10:53:16.22 | -61:35:26.2 | 0.3634 | 53470.017173 | 0.10 | 18.78 |
| 12 | v202 | 10:53:16.83 | -61:30:44.3 | 0.4388 | 53470.218924 | 0.91 | 18.29 |
| 13 | v204 | 10:53:17.08 | -61:34:52.6 | 0.3913 | 53470.153049 | 0.28 | 19.38 |
| 14 | v238 | 10:53:28.14 | -61:26:39.4 | 0.5632 | 53470.183172 | 0.54 | 19.04 |
| 15 | v240 | 10:53:28.41 | -61:35:40.8 | 0.2921 | 53470.246956 | 1.12 | 20.08 |
| 16 | v246* | 10:53:30.70 | -61:31:08.4 | 0.2981 | 53470.977068 | 0.64 | 19.08 |
| 17 | v254 | 10:53:33.02 | -61:22:18.2 | 0.3070 | 53470.062773 | 0.26 | 18.84 |
| 18 | v256 | 10:53:33.82 | -61:24:47.6 | 0.2912 | 53470.089989 | 1.08 | 19.80 |
| 19 | v258 | 10:53:33.85 | -61:33:27.9 | 0.2699 | 53470.170620 | 0.62 | 20.30 |
| 20 | v263 | 10:53:35.53 | -61:25:56.5 | 0.3434 | 53470.006974 | 0.40 | 19.36 |
| 21 | v276 | 10:53:38.14 | -61:31:30.6 | 0.2785 | 53470.111512 | 0.34 | 19.01 |
| 22 | v282 | 10:53:41.74 | -61:24:13.6 | 0.4303 | 53470.149117 | 0.80 | 20.48 |
| 23 | v307 | 10:53:50.19 | -61:32:59.1 | 0.2705 | 53470.002038 | 0.36 | 20.63 |
| 24 | v310 | 10:53:50.55 | -61:32:22.8 | 0.3908 | 53470.155144 | 0.24 | 20.94 |
| 25 | v311* | 10:53:50.59 | -61:24:16.3 | 0.3607 | 53470.185405 | 0.42 | 19.58 |
| 26 | v315 | 10:53:51.68 | -61:23:18.9 | 0.3648 | 53470.218288 | 0.54 | 19.24 |
| 27 | v321 | 10:53:53.38 | -61:29:25.2 | 0.3046 | 53470.309823 | 0.18 | 19.01 |
| 28 | v323* | 10:53:53.74 | -61:26:06.1 | 0.3818 | 53470.112189 | 0.10 | 17.10 |
| 29 | v326 | 10:53:55.06 | -61:31:51.9 | 0.3254 | 53470.243618 | 0.47 | 19.16 |
| 30 | v343 | 10:54:04.79 | -61:23:37.6 | 0.3162 | 53470.195410 | 0.62 | 19.64 |

\* Stars exhibiting O'Connell Effect

## 2. Fourier Decomposition and Classification

It was shown by Rucinski (1993) that the light curves of W UMa type systems can be quantitatively described by using only the two coefficients $a_2$ and $a_4$ of the cosine decomposition $\sum a_i \cos(2\pi i \phi)$. This method has also been utilized in the preliminary classification of 62 eclipsing binary stars selected from the All Sky Automated Survey (ASAS) database which have complementary spectroscopic radial velocity measurements by Deb and Singh (2011). The observed phased light curves of the selected 30 stars were fitted with the Fourier series of the form (24):

$$m(\Phi) = m_0 + \sum_{i=1}^{4}[a_i \cos(2\pi i\Phi) + b_i \sin(2\pi i\Phi)] \quad (1)$$

where $m(\phi)$ denotes the phased light curve. $m_0$ refers to the mean magnitude and the primary minimum ($t_0$) is the zero point of the phased light curve. $t_0$ has been obtained from the three parameter Gaussian fitting to the wings containing the primary minima. Few typical Fourier fits of four stars having star IDs v043, v051, v056 and v186 are displayed in Fig.1. Table 2 presents the Fourier parameters of all the 30 eclipsing binary stars. The distribution of Fourier parameters in the $a_2$-$a_4$ plane are also shown in Fig.2. The continuous envelope to filter out contact binaries as given by the relation $a_4 = a_2(0.125 + a_2)$ (5) is also shown. The stars which lie under this envelope are contact binaries while those above it are semi detached and detached binaries (24, 5).

In our study, almost all the selected eclipsing binary stars were found to be within the continuous envelope and should be classified as contact binaries. However, the classification scheme using the Fourier parameters should be taken to be tentative and preliminary. Final classification scheme of eclipsing binary stars are done using the Roche lobe geometry obtained from the light curve modeling (5).

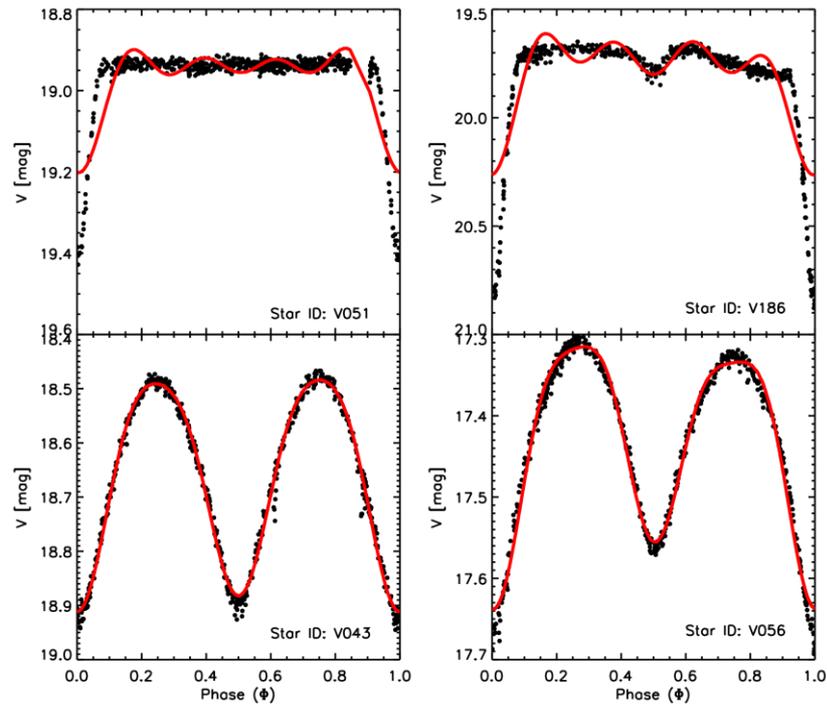

Figure 1: Fourier fitted light curves are shown as solid lines in red color whereas the filled circles in black color denote the original data points.

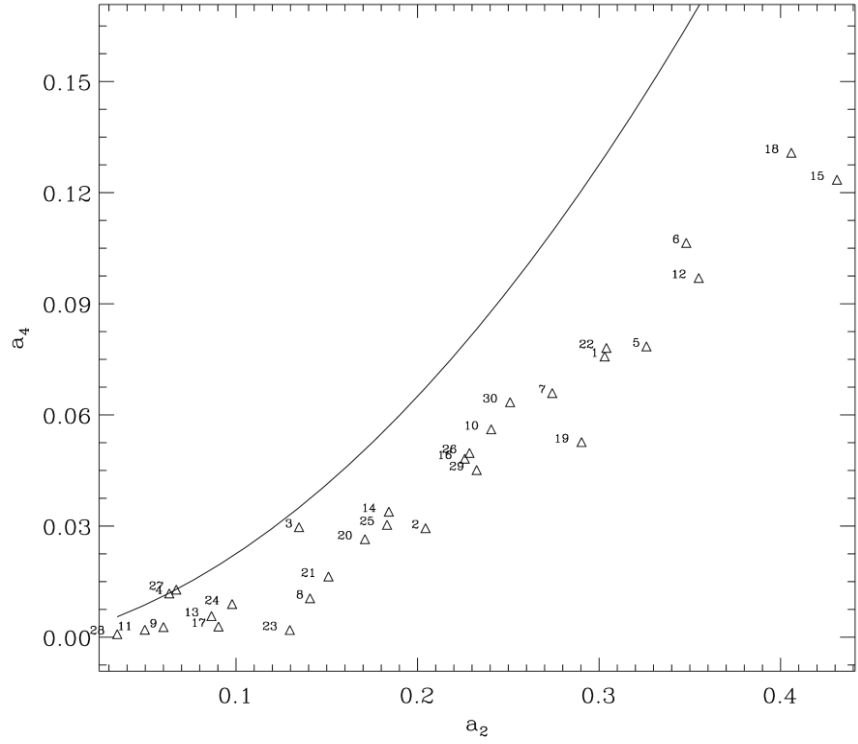

Figure 2: Classification of 30 eclipsing binary stars in the $a_2$-$a_4$ plane of Fourier coefficients. Stars lying below the envelope are contact binaries and are marked by their serial numbers as in Table 2 corresponding to their catalog IDs.

Table 2. A sample of Fourier parameters of all the 30 eclipsing binary stars

| Star ID | $m_0$ | $a_1$ | $b_1$ | $a_2$ | $b_2$ | $a_3$ | $b_3$ | $a_4$ | $b_4$ |
| --- | --- | --- | --- | --- | --- | --- | --- | --- | --- |
| | $\sigma m_0$ | $\sigma a_1$ | $\sigma b_1$ | $\sigma a_2$ | $\sigma b_2$ | $\sigma a_3$ | $\sigma b_3$ | $\sigma a_4$ | $\sigma b_4$ |
| v028 | 18.93954 | 0.05307 | -0.02423 | 0.30307 | 0.02083 | 0.02916 | -0.00625 | 0.07577 | 0.00204 |
| | 0.00041 | 0.00061 | 0.00053 | 0.00059 | 0.00055 | 0.00057 | 0.00056 | 0.00057 | 0.00056 |
| v043 | 18.66279 | 0.00874 | 0.00203 | 0.20441 | -0.00216 | 0.00606 | -0.00154 | 0.02943 | 0.00041 |
| | 0.00033 | 0.00049 | 0.00043 | 0.00047 | 0.00045 | 0.00046 | 0.00046 | 0.00045 | 0.00046 |
| v056 | 17.43158 | 0.03301 | -0.00674 | 0.13471 | 0.01053 | 0.00840 | 0.00131 | 0.02970 | 0.00088 |
| | 0.00016 | 0.00024 | 0.00022 | 0.00022 | 0.00023 | 0.00023 | 0.00022 | 0.00022 | 0.00023 |
| v066 | 17.90496 | 0.00145 | -0.00006 | 0.06323 | 0.00360 | 0.00026 | -0.00010 | 0.01182 | -0.00059 |
| | 0.00021 | 0.00031 | 0.00028 | 0.00031 | 0.00029 | 0.00029 | 0.00030 | 0.00030 | 0.00029 |
| v073 | 19.08748 | 0.03880 | -0.00659 | 0.32609 | 0.00518 | 0.00953 | -0.00377 | 0.07849 | 0.00086 |
| | 0.00049 | 0.00073 | 0.00064 | 0.00070 | 0.00068 | 0.00069 | 0.00067 | 0.00069 | 0.00064 |
| v083 | 21.40271 | 0.05893 | -0.01423 | 0.34801 | 0.02045 | 0.02511 | 0.00155 | 0.10643 | 0.00995 |
| | 0.00257 | 0.00406 | 0.00305 | 0.00391 | 0.00327 | 0.00365 | 0.00330 | 0.00353 | 0.00342 |
| v084 | 19.26458 | 0.03847 | 0.00471 | 0.27420 | 0.00919 | 0.02684 | -0.00163 | 0.06590 | 0.00397 |
| | 0.00048 | 0.00074 | 0.00059 | 0.00071 | 0.00062 | 0.00067 | 0.00064 | 0.00065 | 0.00065 |
| v116 | 19.32783 | 0.01025 | 0.01722 | 0.14074 | -0.01087 | 0.00763 | 0.00064 | 0.01051 | -0.00164 |
| | 0.00052 | 0.00077 | 0.00069 | 0.00074 | 0.00072 | 0.00071 | 0.00073 | 0.00073 | 0.00071 |
| v143 | 18.89983 | 0.00698 | -0.00507 | 0.06004 | 0.01132 | 0.00306 | -0.00056 | 0.00273 | 0.00143 |
| | 0.00037 | 0.00054 | 0.00050 | 0.00052 | 0.00053 | 0.00052 | 0.00052 | 0.00053 | 0.00052 |
| v178 | 20.31959 | 0.06374 | -0.03395 | 0.24055 | -0.00872 | 0.01532 | 0.00098 | 0.05613 | 0.00510 |
| | 0.00108 | 0.00162 | 0.00144 | 0.00157 | 0.00150 | 0.00157 | 0.00148 | 0.00154 | 0.00150 |
| v198 | 18.83179 | 0.00062 | -0.00879 | 0.04976 | -0.00042 | 0.00133 | -0.00110 | 0.00201 | 0.00084 |
| | 0.00037 | 0.00054 | 0.00052 | 0.00053 | 0.00053 | 0.00054 | 0.00052 | 0.00053 | 0.00053 |
| v202 | 18.58668 | 0.08854 | 0.02639 | 0.35488 | -0.01023 | 0.02941 | 0.00285 | 0.09696 | -0.00480 |
| | 0.00036 | 0.00059 | 0.00042 | 0.00054 | 0.00049 | 0.00049 | 0.00048 | 0.00050 | 0.00048 |
| v204 | 19.48034 | 0.07832 | 0.00449 | 0.08647 | 0.01577 | -0.00514 | -0.00557 | 0.00570 | 0.00397 |
| | 0.00057 | 0.00084 | 0.00077 | 0.00082 | 0.00080 | 0.00080 | 0.00081 | 0.00080 | 0.00080 |
| v238 | 19.20603 | -0.05566 | 0.00958 | 0.18418 | -0.00404 | -0.03579 | 0.00321 | 0.03389 | -0.00263 |
| | 0.00045 | 0.00069 | 0.00059 | 0.00065 | 0.00063 | 0.00064 | 0.00062 | 0.00062 | 0.00064 |
| v240 | 20.44112 | 0.04467 | -0.01467 | 0.43104 | 0.01953 | 0.01961 | -0.00606 | 0.12349 | 0.00058 |
| | 0.00119 | 0.00192 | 0.00133 | 0.00182 | 0.00149 | 0.00165 | 0.00148 | 0.00161 | 0.00153 |
| v246 | 19.32125 | 0.04503 | -0.04174 | 0.22600 | 0.00996 | 0.01473 | -0.00329 | 0.04818 | 0.00814 |
| | 0.00049 | 0.00074 | 0.00065 | 0.00070 | 0.00069 | 0.00069 | 0.00069 | 0.00070 | 0.00069 |
| v254 | 19.95462 | 0.01639 | -0.00867 | 0.09034 | 0.00502 | 0.00081 | 0.00014 | 0.00287 | -0.00076 |
| | 0.00075 | 0.00111 | 0.00100 | 0.00103 | 0.00108 | 0.00105 | 0.00105 | 0.00104 | 0.00105 |
| v256 | 20.07273 | 0.04403 | 0.02144 | 0.40591 | -0.01056 | 0.00876 | 0.00040 | 0.13078 | -0.01537 |
| | 0.00083 | 0.00133 | 0.00095 | 0.00126 | 0.00105 | 0.00115 | 0.00106 | 0.00112 | 0.00108 |
| v258 | 20.56386 | 0.01727 | 0.00676 | 0.29034 | -0.02043 | 0.00700 | -0.00283 | 0.05267 | -0.00363 |
| | 0.00121 | 0.00190 | 0.00151 | 0.00176 | 0.00166 | 0.00167 | 0.00167 | 0.00169 | 0.00164 |
| v263 | 19.51942 | 0.00532 | -0.00798 | 0.17101 | -0.01025 | 0.00300 | 0.00087 | 0.02647 | -0.00391 |
| | 0.00056 | 0.00084 | 0.00074 | 0.00081 | 0.00077 | 0.00080 | 0.00078 | 0.00079 | 0.00078 |
| v276 | 19.14913 | 0.01327 | -0.00316 | 0.15092 | 0.00425 | -0.00075 | -0.00265 | 0.01638 | 0.00005 |
| | 0.00044 | 0.00067 | 0.00058 | 0.00063 | 0.00063 | 0.00063 | 0.00061 | 0.00062 | 0.00062 |
| v282 | 20.73390 | 0.08033 | -0.00949 | 0.30404 | -0.02625 | 0.02579 | -0.00225 | 0.07808 | -0.00912 |
| | 0.00134 | 0.00215 | 0.00155 | 0.00201 | 0.00173 | 0.00183 | 0.00174 | 0.00179 | 0.00177 |
| v307 | 20.78366 | -0.02056 | 0.04562 | 0.12964 | -0.03757 | 0.00908 | 0.01217 | 0.00196 | -0.00579 |
| | 0.00142 | 0.00205 | 0.00197 | 0.00203 | 0.00200 | 0.00201 | 0.00201 | 0.00203 | 0.00198 |
| v310 | 21.05112 | -0.00153 | -0.00272 | 0.09785 | 0.00763 | 0.00260 | -0.00225 | 0.00895 | 0.00023 |
| | 0.00176 | 0.00266 | 0.00231 | 0.00253 | 0.00244 | 0.00249 | 0.00244 | 0.00251 | 0.00242 |
| v311 | 19.75348 | 0.00837 | 0.01856 | 0.18318 | 0.00028 | 0.00222 | 0.00146 | 0.03034 | -0.00127 |
| | 0.00062 | 0.00093 | 0.00084 | 0.00088 | 0.00089 | 0.00089 | 0.00087 | 0.00088 | 0.00087 |
| v315 | 19.45422 | 0.00558 | 0.00409 | 0.22861 | 0.01430 | 0.00716 | -0.00076 | 0.04973 | 0.00702 |
| | 0.00052 | 0.00080 | 0.00067 | 0.00077 | 0.00070 | 0.00073 | 0.00073 | 0.00072 | 0.00073 |
| v321 | 19.06996 | 0.01425 | 0.00638 | 0.06707 | -0.00539 | 0.00241 | -0.00097 | 0.01284 | -0.00203 |
| | 0.00048 | 0.00068 | 0.00067 | 0.00066 | 0.00069 | 0.00068 | 0.00066 | 0.00068 | 0.00066 |
| v323 | 17.14975 | 0.00710 | -0.01101 | 0.03456 | 0.00325 | 0.00062 | 0.00163 | 0.00083 | -0.00024 |
| | 0.00014 | 0.00020 | 0.00020 | 0.00020 | 0.00020 | 0.00020 | 0.00020 | 0.00020 | 0.00020 |
| v326 | 19.34989 | 0.00066 | -0.00086 | 0.23256 | 0.00057 | -0.00366 | 0.00247 | 0.04512 | -0.00021 |
| | 0.00053 | 0.00083 | 0.00067 | 0.00077 | 0.00073 | 0.00074 | 0.00073 | 0.00075 | 0.00073 |
| v343 | 19.86415 | 0.04568 | -0.01272 | 0.25097 | -0.00429 | 0.00642 | 0.00346 | 0.06344 | -0.00597 |
| | 0.00073 | 0.00116 | 0.00088 | 0.00108 | 0.00097 | 0.00100 | 0.00099 | 0.00099 | 0.00098 |

### 3. Light curve modeling of eclipsing binary stars

Light curve modeling of eclipsing binary stars was performed using the Wilson-Devinney (WD) code (30,31,32) using PHOEBE (23). Mass ratio ($q$) is the most important parameter in the light curve modeling. It goes as an input in the modeling and should be held fixed. It can be determined accurately only through high resolution spectroscopic radial velocity measurements. Apart from that, the faintness of the sources used in this analysis also places a constrain on the access of such measurements using moderately-sized telescope (2-3 meter class). Since no spectroscopic radial velocity measurements were available for the selected binaries used in this study, $q$ was determined using the extensive $q$-search method from the light curve modeling as discussed in Deb, Singh, Seshadri, and Gupta (2010). In this technique, test solutions were made at the outset for each of the stars. Test solutions were computed at a series of assumed mass-ratios ($q$) with the values ranging from 0.10 to 0.90 in steps of 0.05. Each of the test solutions were computed in mode 3, used for over-contact binaries. For each assumed $q$, a convergent solution was obtained and the resulting sum of the squared deviations $\sum \omega_i (o-c)^2$ of these convergent solutions for each value of $q$ were plotted. The values of $q$ corresponding to the minima of $\sum \omega_i (o-c)^2$ obtained for each of the binary stars were taken as the true mass-ratios to be fixed in the modeling. We ran the code again using these determined mass ratios for each of the stars and let $q$ be adjusted freely along with the other adjustable parameters in order to find out its errors. The mass ratios ($q$) determined from the final solutions are listed in Table. 3. For contact binaries, we had used mode 3 and while for the 1 detached binary found in the sample we have used mode 2 option in the WD code. Throughout the analysis, we have chosen the more massive star as the primary star and label it as star 1 and the less massive star as the secondary star and label it as star 2. Eclipsing of Star 1 and star 2 occur at primary and secondary minimum, respectively. For each of the specific modes in PHOEBE, there are few input parameters that need to be supplied by the user for modeling. The input parameters in mode 3 are the mass-ratio ($q$) star 1 temperature ($T_1$). The bolometric albedos are chosen as: $A_1=A_2=0.5$ for stars having convective envelopes (T < 7200 K) and 1 for radiative envelopes (T > 7200 K). Gravity-darkening coefficients of $g_1=g_2=0.32$ and 1 are chosen for stars having convective and radiative envelopes, respectively, according to Von Zeipel's Law (5). The $q$-search method as discussed above has been used to obtain the mass ratio ($q$) and were held fixed throughout the analysis. Van Hamme's table (29) is used to interpolate the monochromatic and bolometric limb-darkening coefficients. In mode 3, there are two constraints as well. One is the surface potential of star 1 ($\Omega_1$) is equal to that of star 2 ($\Omega_2$) and the other is that the luminosity of star 2 ($L_2$) is determined from other parameters by PHOEBE itself. The adjustable parameters in this mode are: ($T_2$, $i$, $\Omega_1$, $\Omega_2$, $L_1$). Here i is the orbital inclination and $L_1$ represents the monochromatic luminosity of star 1. The computation of luminosity of star 1 is done using a Planck function. In the analysis, the third light parameter and the spot parameter options in the PHOEBE were not used. The inclusion of third light may sometimes affect the orbital inclination and amplitude of the light curves during modeling. The uniqueness of the solution may be affected when the spot parameters are combined for stable solutions (15). This may be prevented by using Doppler imaging techniques (17). Hence these parameter options were avoided during the modeling. Using the mass ratios as determined from the photo metric analysis of the 30 binary stars and their primary temperatures determined from the period-color relation (PCR) as shown in section 3.1, PHOEBE was run for each of the light curves. The fill-out factor or degree of over-contact is given by

$$f = \frac{(\Omega_{in} - \Omega_{1,2})}{\Omega_{in} - \Omega_{out}} \qquad (2)$$

Where $\Omega_{in, out}$ denote the inner, outer Lagrangian surface potentials. For contact binaries $\Omega_1 = \Omega_2 = \Omega$, where $\Omega$ denotes the surface potential of the common envelope of the binary. On the basis of the fill-out factor value or more generally on Roche lobe geometry, one can classify the eclipsing binaries into contact, semi-detached and detached binaries (Deb & Singh, 2011). If $0<f<1$, it belongs to contact binaries. Stars with $f=0$ belong to semi-detached and those with $f<0$ belong to detached binaries (4). In the selected sample of 30 binary stars, only one star comes out to be a detached binary system on the basis of $f$-value. Contact binaries are classified as A-type or W-type accordingly if the primary minimum is a transit or an occultation. Therefore, $M_2/M_1 < 1$ for A-type and $M_2/M_1 > 1$ for W-type contact binaries. On this basis, the type of contact binaries had been mentioned in table 3. Generally the difference in surface temperature between the components ($\Delta T$) < 1000 K for most of the contact binaries. However, if $\Delta T \geq 1000$ K, then those contact binaries are referred to as B-type (3,4). B-type systems were first introduced by Lucy and Wilson (1979). There exists a large surface temperature difference between the components because although being in geometrical contact they are not in contact thermally (3,4). These large differences in temperature cause unequal levels of primary and secondary minima. These binaries exhibit an EB ( Lyrae)-type light curve but their orbital period falls in the range of classical W UMa systems (10, 4).

The WD differential correction (dc) minimization program in PHOEBE yield the values of the fitting parameters and the formal statistical errors as output. Monte Carlo parameter scan (also called heuristic scan) around the best solution was implemented using the PHOEBE-Scripter capabilities (2,4) to investigate the parameter values, errors and stable solutions. The Monte Carlo simulation was run for 1000 times, each time updating the input parameter values to the values determined in the previous iteration of the dc minimization program. Results of the heuristic scan method for the four-parameters fit in the over-contact mode for the star ID v083 are shown as histogram plots in Fig. 3. Errors in the parameters were obtained from the three parameter Gaussian fitting where the resulting distribution is found to be nearly Gaussian.

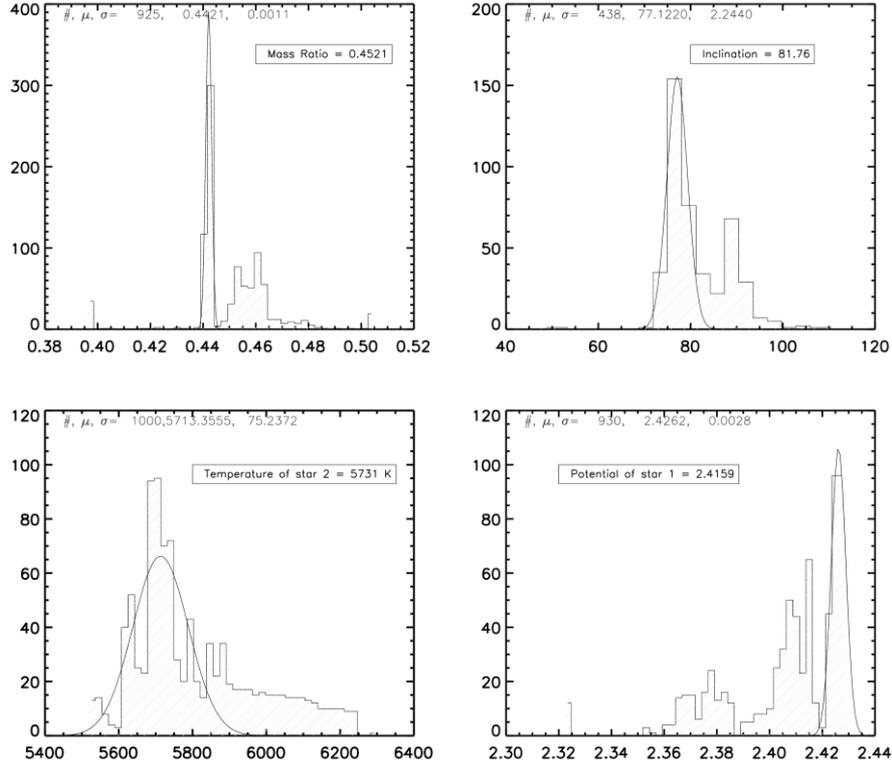

Figure 3: Results of the Monte-carlo parameter scan of star ID v083 for four-parameters $T_2$, $\Omega_i$, $i$ and $q$ of the binary system. The mean and standard deviations of the fitted three parameter Gaussian fit to the region where the distribution approximates nearly Gaussian were taken as the parameter values and their errors, respectively.

Synthetic model light curves of all the 30 stars computed from the WD light curve modeling technique are shown in Figs. 4 5, 6, respectively. The synthetic light curves are shown as solid lines over-plotted on the original phased light curves shown as filled black circles.

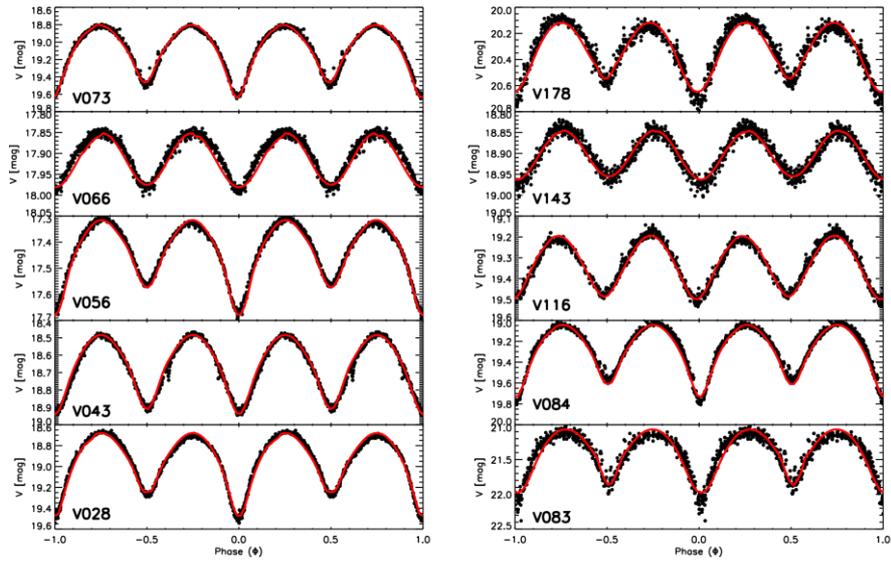

Figure 4: The solid line is the synthetic light curve computed from the WD light-curve modeling technique.

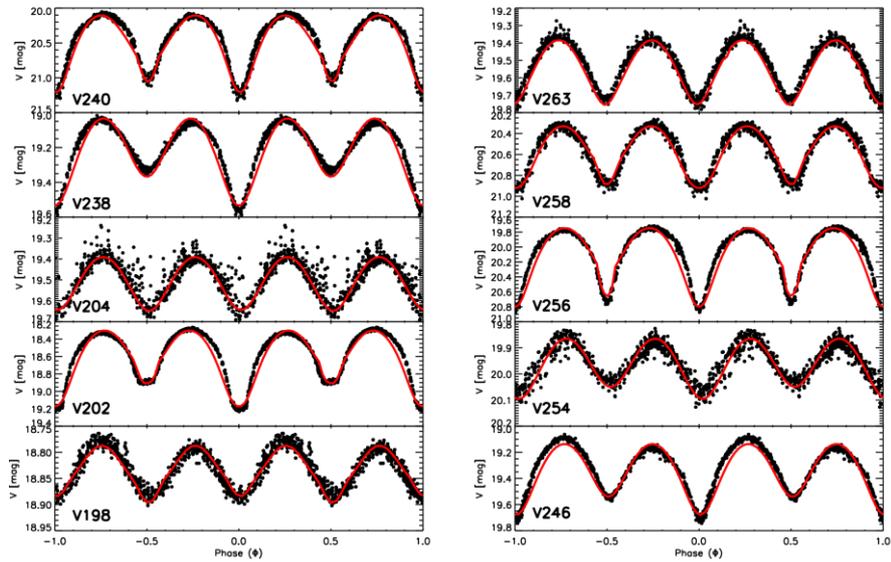

Figure 5: The solid line is the synthetic light curve computed from the WD light-curve modeling technique.

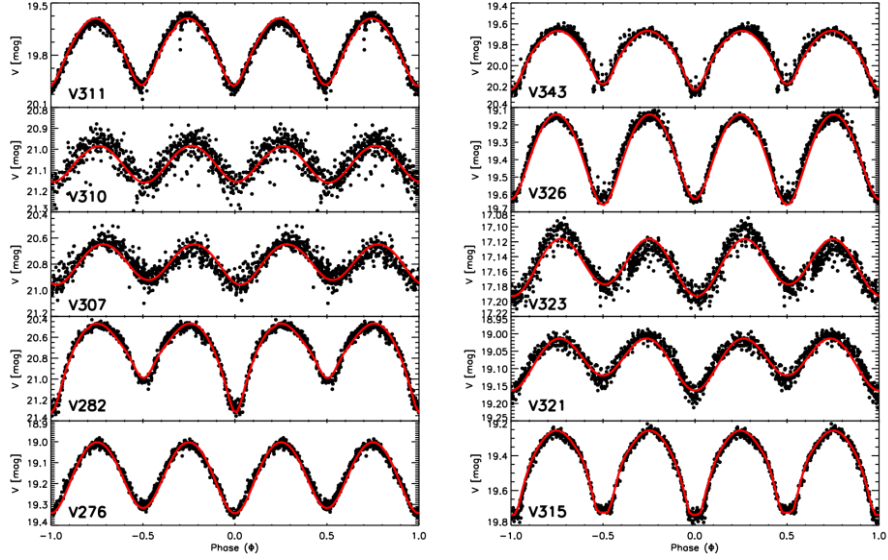

Figure 6: The solid line is the synthetic light curve computed from the WD light-curve modeling technique.

Table 3: Mass ratios of the 30 eclipsing binaries along with their types determined from the light curve modeling. Standard errors are expressed in terms of the last quoted digits.

| Sl.no. | ID | Mass Ratio ($q$) | Type |
|---|---|---|---|
| 1 | v028 | 0.4091(85) | A |
| 2 | v043 | 0.3004(12) | W |
| 3 | v056 | 0.3424(14) | A |
| 4 | v066 | 0.3341(52) | A |
| 5 | v073 | 0.4115(19) | B |
| 6 | v083 | 0.4521(11) | W |
| 7 | v084 | 0.3981(13) | W |
| 8 | v116 | 0.4091(58) | W |
| 9 | v143 | 0.3389(69) | B |
| 10 | v178 | 0.4311(17) | W |
| 11 | v198 | 0.3809(22) | Detached |
| 12 | v202 | 0.4763(51) | A |
| 13 | v204 | 0.4744(39) | W |
| 14 | v238 | 0.5706(30) | B |
| 15 | v240 | 0.5086(07) | W |
| 16 | v246 | 0.4661(22) | A |
| 17 | v254 | 0.4052(50) | A |
| 18 | v256 | 0.4431(14) | A |
| 19 | v258 | 0.4129(83) | A |
| 20 | v263 | 0.3899(96) | A |
| 21 | v276 | 0.3955(51) | A |
| 22 | v282 | 0.4025(27) | B |
| 23 | v307 | 0.3952(65) | W |
| 24 | v310 | 0.3501(69) | W |
| 25 | v311 | 0.4023(17) | A |
| 26 | v315 | 0.2896(33) | W |
| 27 | v321 | 0.4039(81) | W |
| 28 | v323 | 0.4001(42) | A |
| 29 | v326 | 0.3993(19) | W |
| 30 | v343 | 0.3342(37) | A |

*3.1 Temperatures*

In order to determine the effective temperature of the primary component of the binary systems, the period-color relation given by Rucinski (2000) was used. Then the theoretical color-temperature calibrations of Flower (1996) were used to calculate their temperatures. The (*J-K*) color indices were also available for some of the eclipsing binaries in our study in the Two Micron All-Sky Survey (2MASS) Catalogue. The spectral-type of all the 30 eclipsing binaries were determined using the visible and infra-red color indices using the color index-temperature calibration of Flower (1996). The infrared (*J-K*) color from 2MASS data is very useful for spectral classification because of negligible interstellar reddening as compared to (*B-V*) color. $T_2$ was computed using $T_1$ fixed (4). In Table.4, both the color indices along with the determined spectral type for both the components are shown. Temperatures of the primary and secondary component are shown along with other computed parameters using the PHOEBE in Table.6.

Table 4: Period, (*B-V*) and (*J-K*) color index for the 30 eclipsing binary stars. Spectral type of the primary (1) and secondary (2) stars are also listed.

| Sl.no. | ID | Period (d) | (B-V) | (J-K) | Spectral Type (1) | Spectral Type (2) |
|---|---|---|---|---|---|---|
| 1 | v028 | 0.2803 | 0.699694 | 1.2340 | G5 | K1 |
| 2 | v043 | 0.3791 | 0.354702 | - | F2 | F5 |
| 3 | v056 | 0.3348 | 0.469129 | 0.4950 | F5 | G5 |
| 4 | v066 | 0.3104 | 0.556207 | - | F9 | F8 |
| 5 | v073 | 0.4519 | 0.238899 | 0.5240 | A8 | K3 |
| 6 | v083 | 0.2995 | 0.602792 | - | G0 | G2 |
| 7 | v084 | 0.2664 | 0.784527 | 0.3560 | G9 | K1 |
| 8 | v116 | 0.3012 | 0.595164 | 0.3790 | G0 | G1 |
| 9 | v143 | 0.3656 | 0.384854 | 0.2320 | F3 | M4 |
| 10 | v178 | 0.3967 | 0.320273 | 1.1190 | F1 | F5 |
| 11 | v198 | 0.3634 | 0.390116 | - | F3 | F2 |
| 12 | v202 | 0.4388 | 0.255247 | - | A9 | F8 |
| 13 | v204 | 0.3256 | 0.499482 | 0.2040 | G9 | G2 |
| 14 | v238 | 0.5632 | 0.145569 | - | A5 | K2 |
| 15 | v240 | 0.2921 | 0.637696 | - | G2 | G0 |
| 16 | v246 | 0.2981 | 0.609180 | 0.5020 | G0 | F8 |
| 17 | v254 | 0.3070 | 0.570162 | - | F9 | G0 |
| 18 | v256 | 0.2912 | 0.642139 | - | G2 | G1 |
| 19 | v258 | 0.2699 | 0.761822 | 0.2150 | G9 | K3 |
| 20 | v263 | 0.3434 | 0.443108 | 0.8470 | F5 | F3 |
| 21 | v276 | 0.2785 | 0.709910 | - | G6 | G0 |
| 22 | v282 | 0.4303 | 0.266732 | 0.9550 | A9 | G0 |
| 23 | v307 | 0.2705 | 0.758025 | 0.6200 | G8 | G8 |
| 24 | v310 | 0.3908 | 0.331255 | - | F2 | F0 |
| 25 | v311 | 0.3607 | 0.396717 | - | F3 | F3 |
| 26 | v315 | 0.3648 | 0.386756 | 0.5200 | F3 | G0 |
| 27 | v321 | 0.3046 | 0.580320 | - | G0 | G0 |
| 28 | v323 | 0.3818 | 0.349083 | 0.4290 | F2 | G0 |
| 29 | v326 | 0.3254 | 0.500173 | 0.2070 | F7 | F8 |
| 30 | v343 | 0.3162 | 0.533514 | 0.4580 | F8 | G0 |

# RESULTS

Tables 5 and 6 list the geometrical and physical parameters of all the 30 eclipsing binaries obtained in this analysis. The physical parameters were calculated using the parameters obtained in Table. 5. The semi-major axis (*a*) were calculated using the Kepler's third law and the new orbital time periods. These semi-major axes values were used to calculate the radii of the components of each of the 30 binaries using $r = \frac{R}{a}$, where r represents the geometrical mean of the polar, side and back radii of the components. *R* gives the radius of the star. The total mass (*M*) and luminosity (*L*) of all the 30 binary stars were computed using the Maceroni and van't Veer (1996) method. The individual masses ($M_{1,2}$) were calculated using the mass ratio and the total mass (*M*). The luminosity of the components of each binary star were directly calculated using the following equation:

$$\frac{L_j}{L_\odot} = \left(\frac{R_j}{R_\odot}\right)^2 \left(\frac{T_j}{T_\odot}\right)^4 \qquad (3)$$

Here *j* denotes the individual components of each of the binary stars. The geometrical and physical parameter solutions for these 30 eclipsing binaries were obtained for the first time. This would provide an impetus for follow-up high resolution spectroscopic radial velocity measurements for accurate parameter determinations.

Table 5: Results obtained from the light curve modeling.

| Sl.no. | ID | i(°) | ($\Omega_1$) | ($\Omega_2$) | $x_1$ | $x_2$ | $r_1$ | $r_2$ | $T_2(K)$ | $T_1(K)$ | Fill out |
|---|---|---|---|---|---|---|---|---|---|---|---|
| 1 | v028 | 84.88±01.25 | 2.4703±0.001 | 2.4703 | 0.70 | 0.56 | 0.5291 | 0.4012 | 5119±156 | 5560.03 | 0.91 |
| 2 | v043 | 74.69±01.49 | 2.3528±0.013 | 2.3528 | 0.42 | 0.50 | 0.5136 | 0.3646 | 6769±258 | 6941.90 | 0.60 |
| 3 | v056 | 69.54±00.92 | 2.5283±0.086 | 2.5283 | 0.32 | 0.42 | 0.4816 | 0.2950 | 5601± 19 | 6413.16 | 0.11 |
| 4 | v066 | 41.65±00.67 | 2.3986±0.019 | 2.3986 | -0.02 | 0.50 | 0.5368 | 0.3479 | 6298±177 | 6055.03 | 0.81 |
| 5 | v073 | 97.90±01.09 | 2.4625±0.097 | 2.4625 | 0.85 | 0.62 | 0.5273 | 0.3983 | 4780±320 | 7542.09 | 1.01 |
| 6 | v083 | 81.76±02.24 | 2.4159±0.002 | 2.4159 | 0.63 | 0.50 | 0.5250 | 0.4131 | 5731± 75 | 5881.13 | 1.33 |
| 7 | v084 | 98.25±11.32 | 2.4288±0.015 | 2.4288 | 0.38 | 0.50 | 0.5187 | 0.3675 | 5114± 12 | 5321.20 | 0.95 |
| 8 | v116 | 60.12±00.33 | 2.4989±0.022 | 2.4989 | 0.50 | 0.50 | 0.5167 | 0.3789 | 5888±118 | 5908.73 | 0.73 |
| 9 | v143 | 40.42±01.21 | 2.4352±0.091 | 2.4352 | 0.50 | 0.50 | 0.5082 | 0.3262 | 3307± 95 | 6796.72 | 0.56 |
| 10 | v178 | 69.58±02.97 | 2.4435±0.019 | 2.4435 | 0.50 | 0.50 | 0.5336 | 0.3568 | 6662± 37 | 7112.73 | 1.13 |
| 11 | v198 | 50.60±02.56 | 2.6900±0.046 | 2.6900 | 0.50 | 0.50 | 0.4622 | 0.3101 | 7161±287 | 6771.80 | <0 |
| 12 | v202 | 88.78±09.21 | 2.4225±0.102 | 2.4225 | 0.50 | 0.50 | 0.5743 | 0.7025 | 6282± 35 | 7452.32 | 1.39 |
| 13 | v204 | 44.40±17.91 | 2.6070±0.094 | 2.6070 | 0.38 | 0.50 | 0.6549 | 0.6015 | 5820±181 | 5225.49 | 1.08 |
| 14 | v238 | 50.40±18.00 | 2.4900±0.010 | 2.4900 | 0.78 | 0.50 | 0.5291 | 0.4012 | 4900±158 | 5100.07 | 1.54 |
| 15 | v240 | 72.04±00.10 | 2.5192±0.017 | 2.5192 | 1.96 | 0.50 | 0.5310 | 0.3742 | 6034± 12 | 5759.19 | 1.19 |
| 16 | v246 | 60.80±04.00 | 2.4600±0.010 | 2.4600 | 0.50 | 0.50 | 0.5498 | 0.3861 | 6238± 66 | 5858.27 | 1.22 |
| 17 | v254 | 44.80±21.00 | 2.4100±0.008 | 2.4100 | 0.50 | 0.50 | 0.5310 | 0.3852 | 6021±560 | 6001.60 | 1.09 |
| 18 | v256 | 77.20±02.58 | 2.3600±0.020 | 2.3600 | 0.50 | 0.50 | 0.5752 | 0.3942 | 5880± 21 | 5744.18 | 1.49 |
| 19 | v258 | 92.40±00.83 | 2.4200±0.017 | 2.4200 | 0.50 | 0.50 | 0.5408 | 0.5696 | 4871± 8 | 5381.58 | 1.11 |
| 20 | v263 | 62.60±09.51 | 2.4100±0.056 | 2.4100 | 0.50 | 0.50 | 0.5602 | 0.4664 | 6797±632 | 6527.96 | 0.97 |
| 21 | v276 | 55.80±09.67 | 2.4000±0.027 | 2.4000 | 0.50 | 0.50 | 0.5439 | 0.5692 | 6024 | 5529.28 | 1.08 |
| 22 | v282 | 80.03±02.77 | 2.4543±0.019 | 2.4543 | 1.07 | 0.50 | 0.5241 | 0.3823 | 6080± 2 | 7390.45 | 0.91 |
| 23 | v307 | 55.60±34.61 | 2.4200 | 2.4200 | 0.50 | 0.50 | 0.9974 | 0.9734 | 5451±178 | 5391.92 | 1.02 |
| 24 | v310 | 46.80±31.94 | 2.4000±0.063 | 2.4000 | 0.50 | 0.50 | 0.5231 | 0.3527 | 7237± 34 | 7057.64 | 0.80 |
| 25 | v311 | 66.40±17.77 | 2.4560 | 2.4560 | 0.32 | 0.24 | 0.5091 | 0.3582 | 6800±321 | 6740.72 | 0.91 |
| 26 | v315 | 82.80±06.50 | 2.3600±0.084 | 2.3600 | 0.50 | 0.50 | 0.5141 | 0.3043 | 6070±239 | 6787.70 | 0.46 |
| 27 | v321 | 34.40±02.51 | 2.3950±0.062 | 2.3950 | 0.50 | 0.50 | 0.5439 | 0.5692 | 6095 | 5963.43 | 1.16 |
| 28 | v323 | 24.80±11.92 | 2.4000±0.078 | 2.4000 | 0.50 | 0.50 | 0.5395 | 0.6658 | 6030 | 6969.40 | 1.13 |
| 29 | v326 | 92.83±02.98 | 2.4638 | 2.4638 | 0.50 | 0.50 | 0.5548 | 0.5117 | 6177±217 | 6280.73 | 0.07 |
| 30 | v343 | 80.39±21.42 | 2.3357 | 2.3357 | 0.12 | 0.50 | 0.5374 | 0.3673 | 6065±193 | 6144.31 | 0.95 |

Table 6: Physical parameters of the 30 binary stars.

| Sl.no. | ID | $a/R_\odot$ | $M/M_\odot$ | $M_1/M_\odot$ | $M_2/M_\odot$ | $R_1/R_\odot$ | $R_2/R_\odot$ | $L_1/L_\odot$ | $L_2/L_\odot$ |
|---|---|---|---|---|---|---|---|---|---|
| 1 | v028 | 1.909 | 1.192 | 0.846 | 0.346 | 1.010 | 0.766 | 0.875 | 0.361 |
| 2 | v043 | 2.333 | 1.190 | 0.915 | 0.275 | 1.198 | 0.850 | 2.990 | 1.360 |
| 3 | v056 | 2.139 | 1.176 | 0.878 | 0.298 | 1.030 | 0.631 | 1.610 | 0.351 |
| 4 | v066 | 2.034 | 1.176 | 0.884 | 0.292 | 1.092 | 0.708 | 1.438 | 0.707 |
| 5 | v073 | 2.626 | 1.194 | 0.847 | 0.347 | 1.385 | 1.046 | 5.567 | 0.512 |
| 6 | v083 | 2.009 | 1.219 | 0.840 | 0.379 | 1.055 | 0.829 | 1.195 | 0.665 |
| 7 | v084 | 1.823 | 1.149 | 0.827 | 0.322 | 0.945 | 0.669 | 0.642 | 0.275 |
| 8 | v116 | 2.005 | 1.197 | 0.855 | 0.342 | 1.036 | 0.759 | 1.174 | 0.621 |
| 9 | v143 | 2.279 | 1.192 | 0.889 | 0.303 | 1.158 | 0.743 | 2.567 | 0.059 |
| 10 | v178 | 2.463 | 1.279 | 0.894 | 0.385 | 1.314 | 0.879 | 3.965 | 1.365 |
| 11 | v198 | 2.284 | 1.216 | 0.881 | 0.335 | 1.056 | 0.708 | 2.104 | 1.183 |
| 12 | v202 | 2.750 | 1.455 | 0.989 | 0.466 | 1.579 | 1.932 | 6.899 | 5.215 |
| 13 | v204 | 2.140 | 1.245 | 0.847 | 0.398 | 1.401 | 1.287 | 1.313 | 1.705 |
| 14 | v238 | 3.119 | 1.289 | 0.821 | 0.468 | 1.650 | 1.251 | 1.652 | 0.809 |
| 15 | v240 | 1.986 | 1.237 | 0.825 | 0.412 | 1.054 | 0.743 | 1.096 | 0.656 |
| 16 | v246 | 2.008 | 1.227 | 0.840 | 0.387 | 1.104 | 0.775 | 1.288 | 0.816 |
| 17 | v254 | 2.034 | 1.202 | 0.858 | 0.344 | 1.080 | 0.783 | 1.358 | 0.723 |
| 18 | v256 | 1.969 | 1.214 | 0.843 | 0.371 | 1.132 | 0.776 | 1.252 | 0.646 |
| 19 | v258 | 1.855 | 1.180 | 0.837 | 0.343 | 1.003 | 1.057 | 0.508 | 0.840 |
| 20 | v263 | 2.202 | 1.220 | 0.884 | 0.336 | 1.233 | 1.027 | 2.477 | 2.019 |
| 21 | v276 | 1.907 | 1.204 | 0.866 | 0.338 | 1.037 | 1.085 | 0.902 | 1.391 |
| 22 | v282 | 2.534 | 1.184 | 0.846 | 0.338 | 1.328 | 0.969 | 4.720 | 1.151 |
| 23 | v307 | 1.901 | 1.264 | 0.909 | 0.355 | 1.896 | 1.850 | 2.726 | 1.465 |
| 24 | v310 | 2.418 | 1.247 | 0.924 | 0.323 | 1.265 | 0.853 | 3.562 | 1.791 |
| 25 | v311 | 2.280 | 1.227 | 0.876 | 0.351 | 1.160 | 0.817 | 2.492 | 1.280 |
| 26 | v315 | 2.264 | 1.175 | 0.918 | 0.257 | 1.164 | 0.688 | 1.650 | 0.901 |
| 27 | v321 | 2.033 | 1.220 | 0.871 | 0.349 | 1.106 | 1.157 | 1.388 | 1.657 |
| 28 | v323 | 2.429 | 1.324 | 0.946 | 0.378 | 1.310 | 1.617 | 3.632 | 3.101 |
| 29 | v326 | 2.086 | 1.223 | 0.879 | 0.344 | 1.157 | 1.067 | 1.869 | 1.487 |
| 30 | v343 | 2.060 | 1.178 | 0.886 | 0.292 | 1.107 | 0.757 | 1.567 | 0.696 |

*4.1 Distance Determination*

Rucinski and Duerbeck (1997) had derived the following empirical relation with an accuracy of ~0.1 mag to calculate the absolute magnitude of W UMa contact binaries:

$$M_v = -4.44 \log P + 3.02(B-V)_0 + 0.12 \qquad (4)$$

where $P$ represents the orbital period. $(B-V)_0$ represents the intrinsic color index of the star and is given by

$$(B-V)_0 = (B-V) - E(B-V)$$

Here $(B-V)$ is the observed color index and $E(B-V)$ is the interstellar redenning along the line of sight and has been obtained from the Schlegel, Finkbeiner, and Davis (1998) map. Using the above relation, the absolute magnitude of the 30 binary stars were calculated. The apparent magnitudes are the $m_0$ values obtained from the Fourier decomposition method as discussed in section 2. These magnitudes along with their respective absolute magnitudes ($M_v$) are listed in Table.7. Distances of the stars were calculated using the following relation:

$$\frac{D}{10} = 10^{0.2(m_v - M_v + 5 - A_v)}, \qquad (5)$$

where $D$ gives the distance in parsec (pc). In the above equation $m_0$ represents the apparent magnitude ($m_v$), $M_v$ represents the absolute magnitude and $A_v$ denotes the interstellar extinction value along the line of sight for the corresponding equatorial coordinates obtained from the Schlegel et al. (1998) map for the $V$-band. Distances calculated for the 30 binary stars are listed in Table 7.

Table 7: Distance-related parameters for the 30 eclipsing binaries selected in the analysis.

| Sl.No. | Star ID | $M_v$ [mag] | $m_v$ [mag] | $A_v$ [mag] | $D$ [kpc] |
|---|---|---|---|---|---|
| 1 | v028 | 2.177 | 18.93954 | 0.87 | 15.084 |
| 2 | v043 | 0.613 | 18.66279 | 0.82 | 27.923 |
| 3 | v056 | -0.470 | 17.43158 | 1.57 | 18.464 |
| 4 | v066 | -0.635 | 17.90496 | 1.80 | 22.284 |
| 5 | v073 | -0.052 | 19.08748 | 0.83 | 45.909 |
| 6 | v083 | 1.877 | 21.40271 | 0.80 | 55.609 |
| 7 | v084 | 1.523 | 19.26458 | 1.26 | 19.784 |
| 8 | v116 | -0.097 | 19.32783 | 1.53 | 37.929 |
| 9 | v143 | 0.586 | 18.89983 | 0.96 | 29.564 |
| 10 | v178 | 0.565 | 20.31959 | 0.81 | 61.506 |
| 11 | v198 | 0.840 | 18.83179 | 0.77 | 27.820 |
| 12 | v202 | -0.040 | 18.58668 | 0.91 | 34.941 |
| 13 | v204 | 1.660 | 19.48034 | 0.77 | 25.708 |
| 14 | v238 | -1.754 | 19.20603 | 1.23 | 88.309 |
| 15 | v240 | 1.919 | 20.44112 | 0.80 | 35.029 |
| 16 | v246 | 1.696 | 19.32125 | 0.87 | 22.441 |
| 17 | v254 | -0.197 | 19.95462 | 1.45 | 54.995 |
| 18 | v256 | 0.515 | 20.07273 | 1.36 | 43.606 |
| 19 | v258 | 2.416 | 20.56386 | 0.81 | 29.348 |
| 20 | v263 | -0.093 | 19.51942 | 1.30 | 45.971 |
| 21 | v276 | 2.081 | 19.14913 | 0.92 | 16.968 |
| 22 | v282 | -1.636 | 20.73390 | 1.45 | 152.750 |
| 23 | v307 | 2.161 | 20.78366 | 0.90 | 35.037 |
| 24 | v310 | 0.132 | 21.05112 | 0.94 | 99.043 |
| 25 | v311 | -1.196 | 19.75348 | 1.53 | 76.541 |
| 26 | v315 | -1.502 | 19.45422 | 1.55 | 76.075 |
| 27 | v321 | 0.905 | 19.06996 | 1.18 | 24.945 |
| 28 | v323 | -1.125 | 17.14975 | 1.43 | 23.386 |
| 29 | v326 | 0.914 | 19.34989 | 1.00 | 30.703 |
| 30 | v343 | -1.106 | 19.86415 | 1.71 | 71.126 |

CONCLUSION

We have presented a detailed analysis of $V$-band light curves of 30 eclipsing binary stars, mostly contact binaries which were first identified and observed with VIMOS at the Unit Telescope 3 (UT3) of the ESO VLT located at Paranal Observatory from April 9 to 12, 2009 by Pietrukowicz et al. (2009). There were no literature or previous knowledge available about the physical parameters for these binary stars and this is the first attempt to investigate their parameters using the WD code as implemented in the PHOEBE. We had attempted to update the orbital periods as given in Pietrukowicz et al. (2009) using the minimization of entropy technique to get the accuracy of the period but found no substantial improvements. For preliminary classification, we have used cosine decomposition of the light curves into the Fourier coefficients, where contact, semi-detached and detached configurations can be distinguished in the $a_2$-$a_4$ plane. Further, using the Roche lobe geometry, we have confirmed v198 to be a detached system while the rest 29 eclipsing binaries belong to contact binaries. Geometrical and physical parameters of all 30 eclipsing binaries were calculated using the WD light curve modeling technique. Since there were no spectroscopic mass ratios available during the study, photometrically mass ratios were determined from the light curve. Out of 30 eclipsing binary stars, 10 were found to be totally eclipsing binaries and the remaining 20 were partially eclipsing binaries. The parameters determined for the 10 totally eclipsing binaries may be taken to be reliable since the photometric mass ratios obtained from the light curve modeling serves as a substitute for the spectroscopically determined mass ratios which otherwise are very difficult to obtain spectroscopically. The parameters obtained for the partial contact binaries may be taken with caution as there is a degeneracy of the mass ratio found for these configurations (4,25). High-resolution radial velocity measurements obtained from large telescopes like ESO VLT will help in refining the fundamental parameters of these binary systems and hence better understanding of astronomical distance scales.

## ACKNOWLEDGMENTS

The authors gratefully acknowledge the financial assistance from the University of Delhi (DU) through the scheme of DU Innovation Project 2013-14 titled "Astronomy using archival data" (ANDC-203) under whose aegis this work has been completed. The help and support rendered by Prof. H. P. Singh of the Department of Physics & Astrophysics, DU as a mentor of this project is gratefully acknowledged. Thanks are also due to Dr. Savithri Singh, Principal, Acharya Narendra Dev College, New Delhi for making available the necessary infrastructure for successful execution of this project. SD thanks all the collaborators of Indo-US Joint Center on Analysis of Variable Star Data under IUSSTF-DST for their valuable inputs.